\documentclass[12pt]{article}
\usepackage[cp866]{inputenc}
\usepackage{epsfig}

\begin{document}

\title{Elastic scattering and path integral}
\author{Efimov G.V.  \vspace*{0.2\baselineskip}\\
 \itshape LTP JINR,\\
{\it 141980 Dubna, Russia}\vspace*{0.2\baselineskip} }
%
%
\maketitle
\begin{abstract}
Representation of the elastic scattering amplitude in the form of
the path integral is obtained using the stationary Schroedinger
equation. A few methods of evaluation of path integrals for large
coupling constants are formulated. The methods are based on the
uncertainty correlation. The scattering lengths and cross sections
are calculated for the right angled and singular potentials and the
Yukawa potential. The comparison with exact results is made.
\end{abstract}

\section{Introduction}

It is always pleasant to have a solution of a differential equation
in a closed form which permits both to investigate  general
properties of the solution and  perform numerical calculations. In
quantum mechanics, such a closed representation is the solution of
the nonstationary  Schroedinger equation in the form of the Feynman
path or functional integral. In this direction, enormous work is
performed and many monographs and textbooks are published (see, for
example, \cite{Feyn,Klei,Zin,Das} and so on). However, in quantum
mechanics there are plenty important problems with fixed energy,
which require solution of the stationary Schroedinger equation. In
particular, all scattering processes are described by the stationary
Schroedinger equation. At present, there are several approaches (see
\cite{Barb,Chi,Rem,Rosen,Ros}), in which the representation of the
solution of the nonstationary  Schroedinger equation in the path
integral form is used to obtain scattering amplitudes

I would like to remind you what is the difference in the formulation
of the scattering problem in the temporal and stationary quantum
formalism in the case of rapidly decreasing potentials.

Stationary picture. A potential decreases very rapidly so that it
acts in  some bounded region only. Outside of this region particles
can be considered as free ones. Any scattering looks like: at large
distances outside of potential influence there is a current of free
particles in a plane wave form (before scattering) and in a
spherical wave form (after scattering). This physical picture is
really realistic.

Nonstationary picture. It is supposed that in  some infinitely past
$t\to-\infty$ and  some infinitely future $t\to\infty$ the potential
does not exist and all particles are free and are described by plane
waves in all space. The potential is $"$switched on$"$ adiabatically
in  some past and then is $"$switched off$"$ adiabatically in some
future. A  factor $e^{-\epsilon |t|}$ is introduced to describe this
process mathematically. This procedure is called  {\it the
hypothesis of switching on interaction}. This hypothesis is the
basis of the scattering theory in quantum field theory. Physically,
this picture is not correct because any interaction cannot be
switched off. Besides, in quantum field theory space-time regions,
where interaction is absent, do not exist. However, as it turned out
this approach works well. Mathematically, this approach requires an
accurate way to the limiting procedure $\epsilon \to 0$, some rules
should be formulated and some counter terms should be introduced.

Therefore, we consider the nonstationary picture as absolutely
unnecessary to describe scattering in quantum mechanics.

In works \cite{Barb,Chi,Rem,Rosen,Ros}, where the subsequent
references can be found, the path integral method is applied to
potential scattering in nonrelativistic quantum mechanics to get the
representation of the scattering amplitude in the path integral
form. The argumentation is based on the nonstationary picture of
scattering where the hypothesis of switching on interaction plays an
essential role. The temporal Schroedinger equation is considered,
the representation for the $S$-matrix as a transition amplitude from
time $t=-\infty$ to time $t=\infty$ is obtained. The $T$-matrix is
extracted by using the corresponding $\delta$-functions. Besides,
all calculations on this way are quite cumbersome. From my point of
view, this approach is not the best way to get scattering
amplitudes. We will show that the desired representation can be
obtained directly from the stationary Schroedinger equation.

After that the problem arises how to calculate the obtained path
integral. The point is that we can calculate the Gaussian integrals
only. But our path integrals are not the Gaussian type. All known
calculation methods are reduced in any case to appropriate Gaussian
type integrals.

Thus, our problem consists of two points:

1. to get a representation for scattering amplitudes in the form of
the path integral using the stationary Schroedinger equation;

2. to work out comparatively simple methods to evaluate the derived
path integrals.

Our estimations of the path integrals are based on the uncertainty
correlation, which is one of fundamental principles of quantum
mechanics.

It is known that the uncertainty coordinate-time correlation permits
us to evaluate qualitatively and semiquantitatively the spectrum of
any hamiltonian. The main point is very simple. Let a hamiltonian
$H={p^2\over2m}+V(x)$ have a discrete spectrum. Between the middle
size of the region $\Delta x$, where a particle oscillates, and the
middle of its momentum $\Delta p$ the connection does exist $\Delta
p\Delta x\sim \hbar n$, where $n$ is a number of excited states. The
spectrum can be evaluated qualitatively and semiquantitatively by
formula
$$E_n\sim\min\limits_{\Delta x}\left[{\hbar^2n^2\over 2m(\Delta
x)^2}+V(\Delta x)\right].$$

In the scattering case, when particles  belong to the continuous
spectrum, and are free in principle, the phenomenon of the expansion
of the wave packet takes place, i.e. the uncertainty of the
coordinate increases as $(\Delta x)^2\sim t$, when time $t$
increases. We shall see how this property of the wave packet can be
used.

So  our problem is to get simple formulas for qualitative and
semiquantitative estimation of the length of scattering and the
cross section in the case of positive repulsing potentials with a
large coupling constant using the path integral representation of
the scattering amplitude. The Born approximation is not applicable
in these cases. We shall use the results of  \cite{Ef}, where the
representation of the solution of the stationary Schroedinger
equation in the path integral form is obtained.

\section{Elastic scattering amplitude}

The scattering problem is formulated in the following way. A
potential is supposed to be short-range one. The steady-state
Shroedinger equation
\begin{eqnarray*}
&& \left(-{1\over2m}{d^2\over d{\bf x}^2}+V({\bf x})-{{\bf
k}^2\over2m}\right)\Psi({\bf x})=0
\end{eqnarray*}
should be solved for the continuous spectrum $E={k^2\over2m}$ and
the solution should satisfy the asymptotic boundary conditions for
$r\to\infty$
\begin{eqnarray}
\label{A1} && \Psi({\bf x})\longrightarrow e^{i{\bf
kx}}+f(k,\theta){e^{ikr}\over r},
\end{eqnarray}
where $f(k,\theta)$ is the desired scattering amplitude, which
should be found. Let us look for the solution in the form
\begin{eqnarray}
\label{A2} && \Psi({\bf x})=e^{i{\bf kx}}+\Phi({\bf x}).
\end{eqnarray}
The function $\Phi({\bf x})$ satisfies the equation
\begin{eqnarray*}
&& \left(-{1\over2m}{d^2\over d{\bf x}^2}+V({\bf x})-{{\bf
k}^2\over2m}\right)\Phi({\bf x})=-V({\bf x})e^{i{\bf kx}},
\end{eqnarray*}
the solution of which can be written in the form
\begin{eqnarray}
\label{A3} && \Phi({\bf x})=-{1\over -{1\over2m}{d^2\over d{\bf
x}^2}+V({\bf x})-{{\bf k}^2\over2m}-i0}V({\bf x})e^{i{\bf kx}}.
\end{eqnarray}
Let us represent this solution in the path integral form using the
standard calculations  (see \cite{Efim1,Efim2,Ef})
\begin{eqnarray}
\label{A4} && \Phi({\bf x})=-i\int\limits_0^\infty
dt~e^{-it\left[-{1\over2m}{d^2\over d{\bf x}^2}-{{\bf
k}^2\over2m}+V({\bf x})-i0\right]}V({\bf
x})e^{i{\bf kx}}\nonumber\\
&&=-i\int\limits_0^\infty {m^{{3\over2}}dt\over (2\pi
it)^{{3\over2}}}~e^{i{t{\bf k}^2\over2m}}\int d{\bf y}e^{i{{\bf
y}^2m\over2t}}\int{D{\mbox{\boldmath$\xi$}}\over
C}e^{i\int\limits_0^t
d\tau\left[{m\over2}\dot{{\mbox{\boldmath$\xi$}}}^2(\tau)-V\left({\bf
x}+\left(1-{\tau\over t}\right){\bf y}-{\mbox{\boldmath$\xi$}}(\tau)\right)\right]}\nonumber\\
&&\cdot V\left({\bf x}+{\bf y}\right)e^{i{\bf k}\left({\bf x}+{\bf
y}\right)} =-i\int d{\bf y}V\left({\bf y}\right)e^{i{\bf k}{\bf
y}}I({\bf y},{\bf x},{\bf k}),
\end{eqnarray}
where
\begin{eqnarray*}
&&I({\bf y},{\bf x},{\bf
k})\\
&&=\int\limits_0^\infty{m^{{3\over2}}dt\over (2\pi
it)^{{3\over2}}}~e^{{i\over2}\left({t{\bf k}^2\over m}+{({\bf
x}-{\bf y})^2m\over t}\right)}\int{D{\mbox{\boldmath$\xi$}}\over
C}e^{i\int\limits_0^t
d\tau\left[{m\over2}\dot{{\mbox{\boldmath$\xi$}}}^2(\tau)-V\left({{\bf
x}\over t}\tau+\left(1-{\tau\over t}\right){\bf
y}-{\mbox{\boldmath$\xi$}}(\tau)\right)\right]}
\end{eqnarray*}
with the boundary conditions
${\mbox{\boldmath$\xi$}}(0)={\mbox{\boldmath$\xi$}}(t)=0.$ We look
for the behavior of the function $I({\bf y},{\bf x},{\bf k})$ in the
limit $|{\bf x}|=r\to\infty$. Let us introduce new variables $t=rs$
and  ${\bf n}={{\bf x}\over r}$, then for large $r$ one obtains
\begin{eqnarray}
\label{A6} && I({\bf y},{\bf x},{\bf
k})\\
&&={m^{{3\over2}}\over\sqrt{r}}\int\limits_0^\infty {ds\over (2\pi
is)^{{3\over2}}} e^{i{r\over2}\left[{s{\bf k}^2\over m}+{\left({\bf
n}-{{\bf y}\over r}\right)^2m\over
s}\right]}\int{D{\mbox{\boldmath$\xi$}}\over
C}e^{i\int\limits_0^{rs}
d\tau\left[{m\over2}\dot{{\mbox{\boldmath$\xi$}}}^2(\tau)-V\left({{\bf
n}\over s}\tau+\left(1-{\tau\over rs}\right){\bf
y}-{\mbox{\boldmath$\xi$}}(\tau)\right)\right]}\nonumber
\end{eqnarray}
with the boundary conditions
${\mbox{\boldmath$\xi$}}(0)={\mbox{\boldmath$\xi$}}(R)=0$, where
$R=rs\to\infty$.

One essential remark should be made about the boundary condition for
$R\to\infty$. The quadratic form in the path integral measure can
written in the form
\begin{eqnarray*}
&& \int\limits_0^R
d\tau\dot{{\mbox{\boldmath$\xi$}}}^2(\tau)=\int\!\!\!\int\limits_0^R
d\tau
d\tau'{\mbox{\boldmath$\xi$}}(\tau)D^{-1}(\tau,\tau'){\mbox{\boldmath$\xi$}}(\tau')
\end{eqnarray*}
where
$$D^{-1}(\tau,\tau')=-{d^2\over d\tau^2}\delta(\tau-\tau').$$
The Green function of the operator $D^{-1}$, which satisfies zero
boundary conditions, takes the form
$$D(\tau,\tau')=-{1\over2}|\tau-\tau'|+{1\over2}(\tau+\tau')-{\tau\tau'\over
R}$$ This Green function is connected with the solution of the
equation
$$-\ddot{u}(\tau)=J(\tau),~~~~u(0)=u(R)=0.$$
The solution for the finite $R$ is
$$u(\tau)=\int\limits_0^R d\tau'
D(\tau,\tau')J(\tau')=\left(1-{\tau\over R}\right)\int\limits_0^\tau
d\tau'~\tau'J(\tau')+\tau\int\limits_\tau^R
d\tau'\left(1-{\tau'\over R}\right)J(\tau')$$ and satisfies zero
boundary conditions. However for $R\to\infty$ it equals
$$u(\tau)\to\int\limits_0^\tau
d\tau'~\tau'J(\tau')+\tau\int\limits_\tau^\infty d\tau'J(\tau')\to
u(\infty)=\int\limits_0^\infty d\tau'~\tau'J(\tau')\neq0,$$  and the
corresponding Green function is
$$D(\tau,\tau')=-{1\over2}|\tau-\tau'|+{1\over2}(\tau+\tau').$$
Therefore, in the limit $R\to\infty$ we get $u(0)=0$, but
$u(\infty)\neq0$, and, in principle, $u(\infty)$ can be an arbitrary
number.

For large $r$ the integral in (\ref{A6}) over $s$ can be calculated
by the saddle-point method, the result reads
\begin{eqnarray}
\label{A7} && I({\bf y},{\bf x},{\bf k})\to{m~e^{ikr}\over 2\pi
ir}e^{-ik{\bf n}{\bf y}}\int {D{\mbox{\boldmath$\xi$}}\over C}
e^{i\int\limits_0^\infty
d\tau\left[{m\over2}\dot{{\mbox{\boldmath$\xi$}}}^2(\tau)-V\left({{\bf
n}k\over m}\tau+{\bf y}-{\mbox{\boldmath$\xi$}}(\tau)\right)\right]}
\end{eqnarray}
where we have only one condition ${\mbox{\boldmath$\xi$}}(0)=0$.

Thus, the scattering amplitude can be represented as
\begin{eqnarray}
\label{amp1} &&\!\!\!\!\!\!\!\!\!\!\! f(k,\theta)=- {m\over2\pi}\int
d{\bf r}~V\left( r\right)e^{i{\bf qr}+\Phi({\bf r},{\bf k})}
\end{eqnarray}
with
\begin{eqnarray}
\label{amp2} &&e^{\Phi({\bf r},{\bf k})}=\int
{D{\mbox{\boldmath$\xi$}}\over C} e^{i\int\limits_0^\infty
d\tau\left[{m\over2}{\dot{\mbox{\boldmath$\xi$}}}^2(\tau)-
V\left({\bf k}{\tau\over m}+{\bf
r}-{\mbox{\boldmath$\xi$}}(\tau)\right)\right]}\nonumber\\
&&=\int {D{\mbox{\boldmath$\xi$}}\over C}
e^{{i\over2}\int\limits_0^\infty
d\tau{\dot{\mbox{\boldmath$\xi$}}}^2(\tau)}\cdot
e^{-im\int\limits_0^\infty d\tau~V\left({\bf k}\tau+{\bf
r}-{\mbox{\boldmath$\xi$}}(\tau)\right)},~~~~~~
{\mbox{\boldmath$\xi$}}(0)=0.
\end{eqnarray}

The argument in the potential ${{\bf k}\over m}\tau+{\bf
r}-{\mbox{\boldmath$\xi$}}(\tau)$ can be interpreted as a linear
motion ${{\bf k}\over m}\tau+{\bf r}$ plus all possible quantum
fluctuations, which are described by the functional variable
${\mbox{\boldmath$\xi$}}(\tau)$. These quantum fluctuations grow in
time: it is  the phenomenon of the expansion of the wave packet.
This circumstance explains the disappearance of the boundary
condition for $\tau=\infty$.

\subsection{Scattering amplitude and  $"$imaginary$"$ time}

Let us come back to the representation (\ref{A3}). In the case of
positive potentials  ($V({\bf x})>0$) the operator
\begin{eqnarray}
\label{eu1} && -{1\over2m}{d^2\over d{\bf x}^2}+V({\bf x})\geq0
\end{eqnarray}
is positive. Let us introduce the complex variable $z=\kappa+ik$.
The operator
\begin{eqnarray}
\label{eu2} -{1\over2m}{d^2\over d{\bf x}^2}+V({\bf x})+{z^2\over2m}
\end{eqnarray}
on the real part of the positive real axis  $z=\kappa>0$ is
positive. The analytical continuation $z\to-i(k+i0)$ gives the
initial operator in equation  (\ref{A3}). For  $z=\kappa>0$ the
inverse operator can be represented as
\begin{eqnarray}
\label{eu3} && G({\bf x},{\bf y},\kappa)\\
&&={1\over -{1\over2m}{d^2\over d{\bf x}^2}+V({\bf
x})+{\kappa^2\over2m}}\delta({\bf x}-{\bf y})=\int\limits_0^\infty
dse^{-s\left[-{1\over2m}{d^2\over d{\bf x}^2}+V({\bf
x})+{\kappa^2\over2m}\right]}\delta({\bf x}-{\bf y})\nonumber\\
&&=\int\limits_0^\infty{m^{{3\over2}}ds\over (2\pi
s)^{{3\over2}}}~e^{-{1\over2}\left({s\kappa^2\over m}+{({\bf x}-{\bf
y})^2m\over s}\right)}\int{D{\mbox{\boldmath$\xi$}}\over
C}e^{-\int\limits_0^s
d\nu\left[{m\over2}\dot{{\mbox{\boldmath$\xi$}}}^2(\nu)+V\left({{\bf
x}\over s}\nu+\left(1-{\nu\over s}\right){\bf
y}-{\mbox{\boldmath$\xi$}}(\nu)\right)\right]}\nonumber
\end{eqnarray}
with the boundary condition
${\mbox{\boldmath$\xi$}}(0)={\mbox{\boldmath$\xi$}}(s)=0.$

In this representation the variable $s$ has  the meaning of {\it the
imaginary time}.

In order to find the behavior of the function $I({\bf y},{\bf
x},\kappa)$ in the limit $|{\bf x}|=r\to\infty$, we introduce new
variables $s=rv$ and ${\bf n}={{\bf x}\over r}$. Then for large $r$
one obtains
\begin{eqnarray}
\label{eu4} && G({\bf x},{\bf y},\kappa)\\
&&={m^{{3\over2}}\over\sqrt{r}}\int\limits_0^\infty {dv\over (2\pi
v)^{{3\over2}}} e^{-{r\over2}\left[{v{\kappa}^2\over m}+{\left({\bf
n}-{{\bf y}\over r}\right)^2m\over
v}\right]}\int{D{\mbox{\boldmath$\xi$}}\over
C}e^{-\int\limits_0^{rv}
d\nu\left[{m\over2}\dot{{\mbox{\boldmath$\xi$}}}^2(\nu)+V\left({{\bf
n}\over v}\nu+\left(1-{\nu\over rv}\right){\bf
y}-{\mbox{\boldmath$\xi$}}(\nu)\right)\right]}\nonumber\\
&&\to{e^{-\kappa r}\over r}\cdot{m\over 2\pi} e^{\kappa{\bf
ny}}\int{D{\mbox{\boldmath$\xi$}}\over C}e^{-\int\limits_0^R
d\nu\left[{m\over2}\dot{{\mbox{\boldmath$\xi$}}}^2(\nu)+V\left({{\bf
n}\over v}\nu+{\bf
y}-{\mbox{\boldmath$\xi$}}(\nu)\right)\right]}\nonumber
\end{eqnarray}
with the boundary conditions
${\mbox{\boldmath$\xi$}}(0)={\mbox{\boldmath$\xi$}}(R)=0$, where
$R=rv\to\infty$.

Thus, the scattering amplitude in the Euclidean region can be
represented as
\begin{eqnarray}
\label{eu5} &&\!\!\!\!\!\!\!\!\!\!\! F({\bf k},\kappa{\bf n})=
-{m\over2\pi}\int d{\bf y}~V\left( {\bf y}\right)e^{i{\bf
ky}+\kappa{\bf ny}+\Phi({\bf y},\kappa{\bf n})}
\end{eqnarray}
for
\begin{eqnarray}
\label{eu6} &&e^{\Phi({\bf y},\kappa{\bf n})}=\int
{D{\mbox{\boldmath$\xi$}}\over C} e^{-\int\limits_0^\infty
d\nu\left[{1\over2}{\dot{\mbox{\boldmath$\xi$}}}^2(\nu)+
mV\left(\kappa{\bf n}\nu+{\bf
y}-{\mbox{\boldmath$\xi$}}(\nu)\right)\right]},~~~~~~
{\mbox{\boldmath$\xi$}}(0)=0.
\end{eqnarray}
Here the substitution $\nu\to m\nu$ is made.

The analytical continuation $\kappa\to-i(k+i0)$ gives the desired
amplitude for physical momenta. Formally, this representation can be
obtained if in the path integral  (\ref{amp2}) one  goes to
integration over imaginary $"$time$"$ $\tau\to -im\nu$:
\begin{eqnarray}
\label{amp22} &&e^{\Phi({\bf y},{\bf k})}=\int
{D{\mbox{\boldmath$\xi$}}\over C} e^{-\int\limits_0^\infty
ds{1\over2}{\dot{\mbox{\boldmath$\xi$}}}^2(s)- m\int\limits_0^\infty
ds~V\left(-i{\bf k}s+{\bf y}
-{\mbox{\boldmath$\xi$}}(s)\right)},~~~~~~
{\mbox{\boldmath$\xi$}}(0)=0.
\end{eqnarray}
The point is that the physical amplitude is an analytical
continuation  of (\ref{eu6}) for $\kappa\to-i(k+i0)$. But the
question arises, what properties of the potential should provide the
representation  (\ref{amp22}) to be such continuation.

The representation (\ref{eu6}) can be applied directly to scattering
length
\begin{eqnarray}
\label{eu7} &&a= F(0,0,\theta)= -{m\over2\pi}\int d{\bf r}~V\left(
r\right)\int {D{\mbox{\boldmath$\xi$}}\over C}
e^{-\int\limits_0^\infty
d\nu\left[{1\over2}{\dot{\mbox{\boldmath$\xi$}}}^2(\nu)+m
V\left({\bf r}-{\mbox{\boldmath$\xi$}}(\nu)\right)\right]}
\end{eqnarray}
for ${\mbox{\boldmath$\xi$}}(0)=0$.

\subsection{The coordinate system}

Let us choose the following coordinate system:
\begin{eqnarray*}
&&{\bf q}={\bf k}_{in}-{\bf k}_{out},~~~~~~~{\bf v}_{out}={{\bf
k}_{out}\over m}={k\over m}{\bf n},~~~~~{\bf
q}^2=4k^2\sin^2{\theta\over2},\\
&& {\bf k}={\bf k}_{out}=(0,0,k),~~~~~{\bf
q}=\left(0,k\sin\theta,2k\sin^2{\theta\over2}\right),\\
&& {\bf
r}=({\mbox{\boldmath$\rho$}},z)=(\rho\sin\phi,\rho\cos\phi,z).
\end{eqnarray*}
The function $\Phi$ in these coordinates can be written as
\begin{eqnarray}
\label{amp3} &&\Phi({\bf r},{\bf k})=\Phi(\rho,z,k).
\end{eqnarray}
The elastic scattering amplitude has the form
\begin{eqnarray}
\label{amp4} &&\!\!\!\!\!\!\!\!\!\!\! f(k,\theta)=
-m\int\limits_0^\infty d\rho~\rho
J_0\left(k\rho\sin\theta\right)\int\limits_{-\infty}^\infty dz
V\left(\sqrt{\rho^2+z^2}\right)
e^{2ikz\sin^2{\theta\over2}+\Phi(\rho,z,k)},\nonumber\\
\end{eqnarray}
where in the representation (\ref{amp2}) we get ${{\bf k}\tau+\bf
r}=({\mbox{\boldmath$\rho$}},k\tau+z).$

The representation (\ref{amp1}) permits us to get an exact
inequality for the scattering length for any coupling constants. The
scattering length is defined by the integral
\begin{eqnarray}
\label{amp5} &&\!\!\!\!\!\!\!\!\!\!\!
a=f(0,\theta)=-2m\int\limits_0^\infty dr~r^2 V(r)e^{\Phi(r)}
\end{eqnarray}
where
\begin{eqnarray*}
\label{amp6} &&e^{\Phi(r)}=\int {D{\mbox{\boldmath$\xi$}}\over C}
e^{i\int\limits_0^\infty
d\tau\left[{m\over2}{\dot{\mbox{\boldmath$\xi$}}}^2(\tau)-V\left({\bf
r}-{\mbox{\boldmath$\xi$}}(\tau)\right)\right]}=\int
{D{\mbox{\boldmath$\xi$}}\over C} e^{-\int\limits_0^\infty
d\tau\left[{1\over2}{\dot{\mbox{\boldmath$\xi$}}}^2(\tau)+m
V\left({\bf r}-{\mbox{\boldmath$\xi$}}(\tau)\right)\right]}
\end{eqnarray*}
Using the Yensen inequality  one can get
\begin{eqnarray*}
&&e^{\Phi(r)}\geq e^{\Phi_1(r)},\nonumber\\
&&\Phi(r)\geq \Phi_1(r)=-\int {D{\mbox{\boldmath$\xi$}}\over C}
e^{-\int\limits_0^\infty
d\tau{1\over2}{\dot{\mbox{\boldmath$\xi$}}}^2(\tau)}
m\int\limits_0^\infty d\tau V\left({\bf
r}-{\mbox{\boldmath$\xi$}}(\tau)\right)\\
&&=-2m\int{d{\bf p}\over(2\pi)^3}{\tilde{V}(p)\over p^2}e^{i{\bf
pr}}=-2m\left[{1\over r}\int\limits_0^r dy~y^2V(y)+
\int\limits_r^\infty dy~yV(y)\right]
\end{eqnarray*}
and finally
\begin{eqnarray}
\label{amp7} &&\!\!\!\!\!\!\!\!\!\!\! |a|\geq2m\int\limits_0^\infty
dr~r^2 V(r)e^{\Phi_1(r)}.
\end{eqnarray}

Thus, the problem of search for the scattering amplitude is reduced
to calculation of the path integral  (\ref{amp2}) or (\ref{amp22}).
This integral is complicated enough.

One of the effective methods of the path integral evaluation  is the
variational method. However, the variational method based on the
Yensen inequality cannot be applied in our case because the integral
(\ref{amp2}) is complex. The generalization of the variational
method is the method of the Gaussian equivalent representation (see
\cite{Efim1}). The application of this method to the integral
(\ref{amp2}) leads to very cumbersome equations. The solution of
these equations requires many efforts which should to be made for
solution of a  particular task. Therefore, this method will not be
considered in this paper. We refer a reader to \cite{Efim1,Efim2}.

In this paper we formulate method based on the uncertainty
correlation.

So let us consider the integral (\ref{amp2}).

\section{Perturbation theory}

Let us shortly remind perturbation method. The coupling constant
should be small in this case. It means that in the representation
\begin{eqnarray*}
\label{ff} &&e^{\Phi(g)}=\int d\sigma
e^{gW}=\sum\limits_{n=0}^\infty{g^n\over n!}\langle W^n\rangle=
e^{\sum\limits_{n=0}^\infty g^n\Phi_n},\\
&& g\Phi_1=g\langle W\rangle,~~~~~g^2\Phi_2={g^2\over2}\left[\langle
W^2\rangle-\langle W\rangle^2\right],...
\end{eqnarray*}
some few lowest terms should be taken into consideration only.

To perform the calculations of the perturbation terms in
(\ref{amp2}) we will use the standard Gaussian integral  which in
our case has the form
\begin{eqnarray*}
\label{ff} &&\int {D{\mbox{\boldmath$\xi$}}\over C}
e^{{i\over2}\int\limits_0^\infty
d\tau{\dot{\mbox{\boldmath$\xi$}}}^2(\tau)}\cdot
e^{i\int\limits_0^\infty d\tau({\bf
J}(\tau){\mbox{\boldmath$\xi$}}(\tau))}=
e^{-{i\over2}\int\!\!\!\int\limits_0^\infty d\tau d\tau'({\bf
J}(\tau)D(\tau,\tau'){\bf
J}(\tau'))},\\
&&D(\tau,\tau')={1\over2}\left[\tau+\tau'-|\tau-\tau'|\right],~~~~~~D(\tau,\tau)=\tau.
\end{eqnarray*}
The potential in (\ref{amp2}) can be represented as
\begin{eqnarray}
\label{ff} &&V\left(|{\bf k}\tau+{\bf
r}-{\mbox{\boldmath$\xi$}}(\tau)|\right)= \int{d{\bf
p}\over(2\pi)^3}\tilde{V}(p)e^{i{\bf p}({\bf k}\tau+{\bf
r}-{\mbox{\boldmath$\xi$}}(\tau))}
\end{eqnarray}

The formula takes place
\begin{eqnarray*}
\label{ff} &&\int{D{\mbox{\boldmath$\xi$}}\over C}
e^{{i\over2}\int\limits_0^\infty
d\tau{\dot{\mbox{\boldmath$\xi$}}}^2(\tau)}
e^{i\sum\limits_{j=1}^N({\bf p}_j{\mbox{\boldmath$\xi$}}(\tau_j))}
=e^{-{i\over2}\sum\limits_{i,j=1}^N({\bf p}_i{\bf
p}_j)D(\tau_i,\tau_j)}.
\end{eqnarray*}

For $\Phi_1$ and $\Phi_2$ one can get
\begin{eqnarray*}
&& g\Phi_1=-im\int\limits_0^\infty d\tau\int{d{\bf
p}\over(2\pi)^3}\tilde{V}(p)e^{i{\bf p}({\bf k}\tau+{\bf
r})-i{p^2\over 2}\tau} =-2m\int{d{\bf
p}\over(2\pi)^3}{\tilde{V}(p)e^{i(\bf pr)}\over p^2-2({\bf pk})-i0}
\end{eqnarray*}

\begin{eqnarray*}
\label{ff} &&  g^2\Phi_2=-4m^2\int\!\!\!\int{d{\bf p_1}d{\bf
p}_2\over(2\pi)^6}{{\tilde{V}(p_1)\tilde{V}(p_2)e^{i(({\bf p}_1+{\bf
p}_2){\bf r})}}\over(p_1^2-2({\bf
kp}_2)-i0)(p_2^2-2({\bf kp}_2)-i0)}\\
&&\cdot{({\bf p_1p_2})\over(({\bf p}_1+{\bf p}_2)^2-2(({\bf
p}_1+{\bf p}_2){\bf k})-i0)}
\end{eqnarray*}

These functions for $k=0$ define the scattering length and they are
\begin{eqnarray*}
&& g\Phi_1(r)=-2m\left[{1\over r}\int\limits_0^r dy~y^2V(y)+
\int\limits_r^\infty dy~yV(y)\right]
\end{eqnarray*}
and
\begin{eqnarray*}
&&g^2\Phi_2(r)={1\over r}\int\limits_0^r dy~y^2\left({d\over
dy}g\Phi_1(y)\right)^2+ \int\limits_r^\infty dy~y\left({d\over
dy}g\Phi_1(y)\right)^2.
\end{eqnarray*}

All calculations are very simple and we will not consider any
examples.

\section{Linear way approximation and eiconal}

As said above, the argument of the potential in the representation
(\ref{amp2})
$${\bf k}{\tau\over m}+{\bf r}-{\mbox{\boldmath$\xi$}}(\tau)=
{\bf v}\tau+{\bf r}-{\mbox{\boldmath$\xi$}}(\tau)$$ can be
interpreted  as a linear motion of a particle ${\bf v}\tau+{\bf r}$
from the point ${\bf r}$ to infinity plus quantum fluctuations
${\mbox{\boldmath$\xi$}}(\tau)$ around the straight way. These
fluctuations can be evaluated using the uncertainty correlation for
a free motion. In this case, the uncertainty correlations lead to
the following correlations
\begin{eqnarray}
\label{qw0} && \Delta p\Delta r\sim\hbar, ~~~~~~~~\Delta E\Delta
t\sim\hbar,\\
&& E={p^2\over2m},~~~~~ {(\Delta p)^2\over 2m}\Delta t\sim
{\hbar^2\over m(\Delta r)^2}\Delta t\sim \hbar, ~~~~~~~~(\Delta
r)^2\sim {\hbar\over m}\Delta t.\nonumber
\end{eqnarray}
The last correlation $(\Delta r)^2\sim {\hbar\over m}\Delta t$ is
known as extension of the wave packet.

Let us come back to the path integral
\begin{eqnarray}
\label{qw1} &&e^{\Phi(\rho,z,k)}=\int {D{\mbox{\boldmath$\xi$}}\over
C} e^{i\int\limits_0^\infty
d\tau\left[{m\over2}{\dot{\mbox{\boldmath$\xi$}}}^2(\tau)-V\left({\bf
k}{\tau\over m}+{\bf
r}-{\mbox{\boldmath$\xi$}}(\tau)\right)\right]}.\nonumber
\end{eqnarray}
The main contribution to this integral comes from
$|{\mbox{\boldmath$\xi$}}(\tau)|\sim\sqrt{\hbar{\tau\over m}}$.
Therefore, in order to neglect the function
${\mbox{\boldmath$\xi$}}(\tau)$ in the potential, the next condition
should be satisfied
$${|{\bf k}|\over m}\tau\gg|{\mbox{\boldmath$\xi$}}(\tau)|\sim\sqrt{\hbar{\tau\over m}}.$$
Let $r_0$ be of an order of the size of potential action. Then into
integral over $\tau$ the main contribution comes from the regions
$$k{\tau\over m}\sim r_0,~~~~~~~~~|{\mbox{\boldmath$\xi$}}(\tau)|\sim\sqrt{\hbar{\tau\over
m}}\sim r_0.$$ These correlations show that for large momenta  $k$
when  $k\gg{\hbar\over r_0}$ one can neglect quantum fluctuations in
the potential, i.e.
$$V\left({\bf k}{\tau\over m}+{\bf
r}-{\mbox{\boldmath$\xi$}}(\tau)\right)\approx V\left({\bf
k}{\tau\over m}+{\bf r}\right).$$

As a result, we get the so called  linear way or eiconal
approximation:
\begin{eqnarray}
\label{qw3} &&e^{\Phi(\rho,z,k)}\approx e^{-i\int\limits_0^\infty
d\tau V\left({\bf k}{\tau\over m}+{\bf r}\right)}=e^{-{im\over
k}\int\limits_z^\infty ds~ V\left(\sqrt{s^2+\rho^2}\right)}.
\end{eqnarray}
It should be noted that the function $\Phi(\rho,z,k)$ is diverged
for $k=0$, i.e. the eiconal approximation is valid for large momenta
only.

The elastic scattering amplitude takes the form
\begin{eqnarray}
\label{qww4} && f(k,\theta)\\
&&\approx -m\int\limits_0^\infty d\rho~\rho
J_0\left(k\rho\sin\theta\right)\int\limits_{-\infty}^\infty dz
V\left(\sqrt{\rho^2+z^2}\right) e^{2ikz\sin^2{\theta\over2}-{im\over
k}\int\limits_z^\infty ds~
V\left(\sqrt{s^2+\rho^2}\right)}.\nonumber
\end{eqnarray}

For small scattering angles $\theta$ the representation (\ref{qww4})
turns into the well-known quasiclassical approximation. Really, for
large momenta and small angles one can estimate
$$k\theta\sim1,~~~~~\theta\sim{1\over k}\ll1,$$
so that in the representation  (\ref{qww4}) one can put
$$e^{2ikz\sin^2\left({\theta\over2}\right)}\approx1.$$
As a result, the well known eiconal approximatin arises
\begin{eqnarray}
\label{qw5} &&\!\!\!\!\!\!\!\!\!\!\! f(k,\theta)\approx
-m\int\limits_0^\infty d\rho~\rho
J_0\left(k\rho\theta\right)\int\limits_{-\infty}^\infty dz
V\left(\sqrt{\rho^2+z^2}\right) e^{-{im\over k}\int\limits_z^\infty
ds~ V\left(\sqrt{s^2+\rho^2}\right)}\nonumber\\
&& =ik\int\limits_0^\infty d\rho~\rho J_0\left(\rho
k\theta\right)\left[1- e^{-{im\over k}\int\limits_{-\infty}^\infty
ds~ V\left(\sqrt{s^2+\rho^2}\right)}\right].
\end{eqnarray}
This amplitude leads to the cross section
\begin{eqnarray}
\label{qw6} && \sigma(k)=2\pi\int\limits_0^\pi
d\theta\sin(\theta)|f(k,\theta)|^2= 2\pi\int\limits_0^\infty
d\rho~\rho \left|1- e^{-i{m\over k}\int\limits_{-\infty}^\infty
d\tau V\left(\sqrt{\tau^2+\rho^2}\right)}\right|^2\nonumber\\
&&=8\pi\int\limits_0^\infty d\rho~\rho \sin^2\left({m\over
2k}\int\limits_{-\infty}^\infty d\tau
V\left(\sqrt{\tau^2+\rho^2}\right)\right).
\end{eqnarray}

It should be noted that formula (\ref{qww4})can be considered as
generalization of the standard quasiclassical approximation for all
angles.

For large momenta one gets
\begin{eqnarray}
\label{qw7} &&{\rm Im} f(k,0)=k\int\limits_0^\infty d\rho~\rho
\left[1- \cos\left({m\over
k}Y(\rho)\right)\right]\approx{m^2\over2k}\int\limits_0^\infty
d\rho~\rho Y^2(\rho),
\end{eqnarray}
where
$$Y(\rho)=\int\limits_{-\infty}^\infty
ds~V\left(\sqrt{s^2+\rho^2}\right).$$

The cross section for large momenta equals
\begin{eqnarray}
\label{qw8} &&\sigma(k)=2\pi{m^2\over k^2}\int\limits_0^\infty
d\rho~\rho Y^2(\rho)={4\pi\over k}{\rm Im} f(k,0)
\end{eqnarray}
in complete correspondence with the unitary requirement.

For small and intermediate momenta the eiconal approximation does
not work. As the next step in taking into account quantum
fluctuations we formulate two approaches which will be named  {\it
the quantum mean approximation} and {\it the unitary approximation}.

\section{Quantum mean approximation}

The linear way approximation  supposes that all quantum fluctuations
can be neglected. This approximation is justified for large momenta.
If the momenta are small, quantum fluctuations play an essential
role and should be taken into account. The natural generalization of
the eiconal formulas will be suggested the following approach.

Let us consider the path integral (\ref{eu7}), describing the
scattering length. In this case, momenta equal zero $k=0$, and
quantum correlations are important. This integral is real and the
main contribution to it comes from the region of the order
\begin{eqnarray}
\label{pqs1}
&&\langle{\mbox{\boldmath$\xi$}}(\nu)\rangle^2_{{\mbox{\boldmath$\xi$}}}=
\int {D{\mbox{\boldmath$\xi$}}\over C} e^{-\int\limits_0^\infty
d\nu{1\over2}{\dot{\mbox{\boldmath$\xi$}}}^2(\nu)}
{\mbox{\boldmath$\xi$}}^2(\nu)\sim\nu
\end{eqnarray}
It is in correspondence with  the uncertainty correlation
(\ref{qw0}), i.e. quantum fluctuations increase as
$$\langle{\mbox{\boldmath$\xi$}}(\nu)\rangle^2_{{\mbox{\boldmath$\xi$}}}\sim \nu.$$

{\it The quantum mean approximation} is a way to take into account
these fluctuations. We suppose that the main contribution to the
path integral comes from the region (\ref{pqs1}), and mathematically
it is realized by formula
\begin{eqnarray}
\label{pqs2} &&e^{\Phi(r)}=\int {D{\mbox{\boldmath$\xi$}}\over C}
e^{-\int\limits_0^\infty
d\nu\left[{1\over2}{\dot{\mbox{\boldmath$\xi$}}}^2(\nu)+mV\left({\bf
r}-{\mbox{\boldmath$\xi$}}(\nu)\right)\right]}\nonumber\\
&&\approx e^{-m\int\limits_0^\infty d\nu V\left(\sqrt{
r^2+\langle{\mbox{\boldmath$\xi$}}(\nu)\rangle^2}\right)}=
e^{-m\int\limits_0^\infty d\nu V\left(\sqrt{ r^2+b\nu}\right)}.
\end{eqnarray}
The value of the parameter  $b\sim1$ is of an order of unity and the
variation $b=1\pm\delta$ defines the accuracy of this approximation.
So that we get
\begin{eqnarray}\label{pqs3}
&&e^{\Phi(r)}\approx e^{-m\int\limits_0^\infty d\tau V\left(\sqrt{
r^2+b\tau}\right)} =e^{-{2m\over b}\int\limits_r^\infty ds~s
V\left(s\right)}.
\end{eqnarray}

The representation for the scattering length reads
\begin{eqnarray}
\label{pqs4} &&a= 2m\int\limits_0^\infty dr~r^2V\left(
r\right)e^{-{2m\over b}\int\limits_r^\infty ds~s V\left(s\right)}.
\end{eqnarray}
The cross section  $\sigma(k)$ for zero momentum $k=0$ equals
\begin{eqnarray}
\label{pqs5} &&\!\!\!\!\!\!\!\!\!\!\!\sigma(0)=4\pi a^2=
4\pi\left[2m\int\limits_0^\infty dr~r^2V\left( r\right)e^{-{2m\over
b}\int\limits_r^\infty ds~s V\left(s\right)}\right]^2.
\end{eqnarray}

\vspace{1cm}

In order to get the cross section for all momenta, we proceed in the
following way. We know the behavior of the cross section for large
(\ref{qw6}) and zero (\ref{pqs5}) momenta. Therefore, we need to
connect these sections in a smooth way. To do this we suggest two
approaches which will be named  {\it the quantum mean approximation}
and {\it unitary approximation}.

\vspace{1cm}

{\it The quantum mean approximation for the amplitude} consists in
that the scattering amplitude is represented in the form
\begin{eqnarray}
\label{pqs6} &&f(k,\theta)\\
&&= -m\int\limits_0^\infty d\rho~\rho
J_0\left(k\rho\sin\theta\right)\int\limits_{-\infty}^\infty dz
V\left(\sqrt{\rho^2+z^2}\right) e^{2ikz\sin^2{\theta\over2}-{m\over
k_c-ik}\int\limits_z^\infty ds
V\left(\sqrt{s^2+\rho^2}\right)},\nonumber
\end{eqnarray}
where the parameter $k_c$ is introduced. This representation does
not change high momentum behavior. The value of the parameter $k_c$
is defined by the condition that for zero momenta the amplitude
$f(0,\theta)$ coincides with the scattering length (\ref{pqs4}):
\begin{eqnarray}
\label{pqs7} &&\!\!\!\!\!\!\!\!\!\!\!  2m\int\limits_0^\infty
dr~r^2V\left( r\right)e^{-{2m\over b}\int\limits_r^\infty ds~s
V\left(s\right)}= k_c\int\limits_0^\infty d\rho~\rho \left[1-
e^{-{m\over k_c}\int\limits_{-\infty}^\infty ds
V\left(\sqrt{s^2+\rho^2}\right)}\right].
\end{eqnarray}
This equation defines the parameter  $k_c$. Then the cross section
equals
\begin{eqnarray}
\label{pqs8} &&\sigma(k)=2\pi \int\limits_0^\pi
d\theta\sin\theta|f(k,\theta)|^2.
\end{eqnarray}

\vspace{1cm}

{\it The quantum mean approximation for the cross section} consists
in modification of the eiconal formula (\ref{qw6}):
\begin{eqnarray}
\label{pqs9} &&\sigma_a(k)\approx 8\pi \int\limits_0^\infty
d\rho~\rho \sin^2\left({m\over\sqrt{k_c^2+k^2}}\int\limits_0^\infty
dz V\left(\sqrt{\rho^2+z^2}\right)\right).
\end{eqnarray}
The parameter $k_c$ is defined by the condition that for the zeroth
momenta the cross sections in two formulas   (\ref{pqs5}) and
(\ref{pqs9}) coincide, i.e.
\begin{eqnarray}
\label{pqs7} && \sigma_0(b)=\sigma_a(0,k_c),\\
&&4\pi\left[2m\int\limits_0^\infty dr~r^2V\left(
r\right)e^{-{2m\over b}\int\limits_r^\infty ds~s
V\left(s\right)}\right]^2=8\pi \int\limits_0^\infty d\rho~\rho
\sin^2\left({m\over k_c}\int\limits_0^\infty dz
V\left(\sqrt{\rho^2+z^2}\right)\right).\nonumber
\end{eqnarray}
It is the equation to calculate the parameter $k_c$.

The accuracy of its approximation is controlled by the parameter $b$
which is changed in the vicinity of the point $b\sim1$.

It should be noted that these approximations are rough enough, but
the general character of the behavior is described correctly.

Another approach of  {\it the quantum mean approximation } can be
the following approximation for the path integral
\begin{eqnarray}
\label{pqs10} &&e^{\Phi({\bf k},{\bf r})}=\int
{D{\mbox{\boldmath$\xi$}}\over C} e^{i\int\limits_0^\infty
d\tau\left[{m\over2}{\dot{\mbox{\boldmath$\xi$}}}^2(\tau)-V\left({\bf
k}{\tau\over m}+{\bf
r}-{\mbox{\boldmath$\xi$}}(\tau)\right)\right]}\\
&& \approx \int{d{\mbox{\boldmath$\xi$}}\over(2\pi i)^{{3\over2}}}
e^{{i\over2}{\mbox{\boldmath$\xi$}}^2-im\int\limits_0^\infty d\nu
V\left({\bf k}\nu+{\bf
r}-b{\mbox{\boldmath$\xi$}}\sqrt{\nu}\right)}.\nonumber
\end{eqnarray}
Further work  should be done to evaluate the effectiveness of this
approach.

\newpage
\section{Unitary approximation}

The cross section for small momenta can be obtained in  another way
which is named {\it the unitary approximation}. The scattering
amplitude is supposed to have the form
\begin{eqnarray}
\label{up1} &&f(k,\theta)\\
&&= -m\int\limits_0^\infty d\rho~\rho
J_0\left(k\rho\sin\theta\right)\int\limits_{-\infty}^\infty dz
V\left(\sqrt{\rho^2+z^2}\right) e^{2ikz\sin^2{\theta\over2}-{m\over
k_c-ik}\int\limits_z^\infty ds
V\left(\sqrt{s^2+\rho^2}\right)},\nonumber
\end{eqnarray}
However, the parameter  $k_c$ and the cross section will be defined
by the unitary condition.

The amplitude for the forward scattering $\theta=0$ equals
\begin{eqnarray}
\label{up2} &&f(k,0)= -m\int\limits_0^\infty d\rho~\rho
\int\limits_{-\infty}^\infty dz V\left(\sqrt{\rho^2+z^2}\right)
e^{-{m\over
k_c-ik}\int\limits_z^\infty ds V\left(\sqrt{s^2+\rho^2}\right)}\nonumber\\
&&=(-k_c+ik)\int\limits_0^\infty d\rho~\rho\left[1- e^{-{m\over
k_c-ik}Y(\rho)}\right],\\
&& Y(\rho)=\int\limits_{-\infty}^\infty ds
V\left(\sqrt{s^2+\rho^2}\right).\nonumber
\end{eqnarray}

The imaginary part of the amplitude reads
\begin{eqnarray}
\label{up3}&&{{\rm Im}f(k,0)\over k}\\
&&=\int\limits_0^\infty d\rho~\rho\left\{1- e^{-{mk_c\over
k_c^2+k^2}Y(\rho)}\left[\cos\left({mk\over k_c^2+k^2}Y(\rho)\right)+
{k_c\over k}\sin\left({mk\over
k_c^2+k^2}Y(\rho)\right)\right]\right\},\nonumber
\end{eqnarray}

The unitary condition, or the optical theorem requires
\begin{eqnarray}
\label{up4} &&\sigma(k)={4\pi\over k}{\rm
Im}(k,0)=2\pi\int\limits_0^\pi d\theta \sin\theta |f(k,\theta)|^2.
\end{eqnarray}

{\it The unitary approximation} consists in that the parameter $k_c$
is a function of momentum $k_c=k_c(k)$, and the unitary condition
(\ref{up4}) is the equation on this function.

Now we proceed in a simpler way. For the zeroth momentum ($k=0$) the
imaginary part of the scattering amplitude reads
\begin{eqnarray*}
&&\left.{{\rm Im}f(k,0)\over k}\right|_{k=0} =\int\limits_0^\infty
d\rho~\rho\left[1- e^{-{m\over k_c}Y(\rho)}\left(1+{m\over
k_c}Y(\rho)\right)\right].
\end{eqnarray*}
and, according to the unitary condition, the cross section equals
\begin{eqnarray}
\label{up6} &&\sigma(0)=4\pi\left.{{\rm Im}f(k,0)\over
k}\right|_{k=0}=4\pi A^2(k_c),
\end{eqnarray}
where $A(k_c)$ is the scattering length:
\begin{eqnarray*}
\label{up7} &&A(k_c)=f(0,0)=-k_c\int\limits_0^\infty
d\rho~\rho\left[1- e^{-{m\over k_c}Y(\rho)}\right].
\end{eqnarray*}

The correlation (\ref{up6}) leads to the equation for the parameter
$k_c$:
\begin{eqnarray}
\label{up8} &&4\pi\left(k_c\int\limits_0^\infty d\rho~\rho\left[1-
e^{-{m\over k_c}Y(\rho)}\right]\right)^2=4\pi\int\limits_0^\infty
d\rho~\rho\left[1- e^{-{m\over k_c}Y(\rho)}\left(1+{m\over
k_c}Y(\rho)\right)\right].\nonumber\\
\end{eqnarray}
Finally, the cross section looks like
\begin{eqnarray}
\label{up9} &&\sigma(k)={4\pi\over k}{\rm Im}(k,0)\\
&&=4\pi\int\limits_0^\infty d\rho~\rho\left\{1- e^{-{k_c~
mY(\rho)\over k_c^2+k^2}}\left[\cos\left({k~mY(\rho)\over
k_c^2+k^2}\right)+ {k_c\over k}\sin\left({k~mY(\rho)\over
k_c^2+k^2}Y\right)\right]\right\}.\nonumber
\end{eqnarray}

Below we will test the effectiveness of these approximations.

\section{Examples}

Let us consider some examples to demonstrate  the approaches
formulated above. We calculate the scattering lengths and cross
sections for the right angled and singular potentials and the Yukawa
potential and compare these approximations with exact results.

\subsection{Right angled potential}

Let us consider the right angled repulsing potential
\begin{eqnarray*}
\label{ff} && V(r)=V\theta(R-r)=\left\{\begin{array}{cc}
V,& r<R;\\
0, & R<r.\\
\end{array}\right.
\end{eqnarray*}
The exact solution can be obtained in the standard way. The wave
function is represented as a sum over partial waves:
\begin{eqnarray*}
\label{ff} && \Psi({\bf r})=\sum\limits_{\ell=0}^\infty
{\chi_\ell(r)\over\sqrt{r}}Y_{\ell~m}({\bf n}).
\end{eqnarray*}
The solution of the Schroedinger equation looks like
\begin{eqnarray*}
\label{ff} && \chi_\ell(r)=\left\{\begin{array}{cc}
C_\ell I_{\ell+{1\over2}}(r\kappa), & r<R;\\
A_\ell(k)J_{\ell+{1\over2}}(kr)+B_\ell(k)N_{\ell+{1\over2}}(kr), & R<r.\\
\end{array}\right.
\end{eqnarray*}
Here the notions  $\kappa=\sqrt{G-k^2},~G=2mV$ are introduced. For
large $r$ we have
\begin{eqnarray*}
\label{ff} &&\chi_\ell(r)=
A_\ell(k)J_{\ell+{1\over2}}(kr)+B_\ell(k)N_{\ell+{1\over2}}(kr)\\
&&\to \sqrt{{2\over\pi r}}\sqrt{A_\ell^2(k)+B_\ell^2(k)}
\sin\left(kr-{\pi\over2}\ell+\delta_\ell(k)\right),~~~r\to\infty.
\end{eqnarray*}
The boundary conditions for $r=R$
\begin{eqnarray*}
\label{ff} &&
C_\ell I_{\ell+{1\over2}}(\kappa R)=A_\ell(k)J_{\ell+{1\over2}}(kR)+B_\ell(k)N_{\ell+{1\over2}}(kR)\\
&&\kappa C_\ell I_{\ell+{1\over2}}'(\kappa
R)=k\left[A_\ell(k)J_{\ell+{1\over2}}'(kR)+B_\ell(k)N_{\ell+{1\over2}}'(kR)\right]
\end{eqnarray*}
define the phases
\begin{eqnarray*}
&&\tan\delta_\ell(k)={B_\ell(k)\over A_\ell(k)}=-{k
I_{\ell+{1\over2}}(\kappa R)J_{\ell+{1\over2}}'(kR)-\kappa
I_{\ell+{1\over2}}'(\kappa R)J_{\ell+{1\over2}}(kR) \over k
I_{\ell+{1\over2}}(\kappa R)N_{\ell+{1\over2}}'(kR)-\kappa
I_{\ell+{1\over2}}'(\kappa R)N_{\ell+{1\over2}}(kR)}.
\end{eqnarray*}
The scattering amplitude has the standard form
\begin{eqnarray*}
\label{ff} && f(k,\theta)={1\over
2ik}\sum\limits_{\ell=0}^\infty(2\ell+1)\left(e^{2i\delta_\ell(k)}-1\right)P_\ell(\cos\theta)\\
\end{eqnarray*}
The scattering length equals
\begin{eqnarray*}
\label{ff} && a(G)=f(0,0)=\left.{\delta_\ell(k)\over
k}\right|_{k=0}=R\left[1-{\tanh(\sqrt{G})\over\sqrt{G}}\right]
\end{eqnarray*}
The cross section is
\begin{eqnarray*}
\label{ff} && \sigma(k)={4\pi\over k^2}
\sum\limits_{\ell=0}^\infty(2\ell+1)\sin^2(\delta_\ell(k)).
\end{eqnarray*}

\subsubsection{Quantum mean approximation}

In the case of the right angled repulsing potential we get
\begin{eqnarray*}
\label{ff}
&&2mV\left(\sqrt{\rho^2+s^2}\right)=G\theta\left(R-\sqrt{\rho^2+s^2}\right)=G\theta\left(\sqrt{R^2-\rho^2}-s\right),\\
&&2m\int\limits_r^\infty ds~sV(s)=2mV\int\limits_r^R
ds~s={G\over2}(R^2-r^2),\\
&&mY(\rho)=2m\int\limits_0^\infty ds~
V\left(\sqrt{s^2+\rho^2}\right)=G\sqrt{R^2-\rho^2}.
\end{eqnarray*}
In the quantum mean approximation the scattering length equals
\begin{eqnarray*}
\label{ff} && a=G\int\limits_0^R dr~r^2 e^{-{G\over
2b}\left(R^2-r^2\right)}=bR^2\left[1-{1\over2}\sqrt{{\pi\over
c}}e^{-c}{\rm Erfi}(\sqrt{c})\right],~~~c={GR^2\over b}
\end{eqnarray*}

The scattering amplitude in this approximation takes the form
\begin{eqnarray*}
\label{qw4} &&f(k,\theta,k_a)\\
&&={G\over2i}\int\limits_0^R d\rho~\rho
J_0\left(k\rho\sin\theta\right){ \left[
e^{2ik\sin^2{\theta\over2}\sqrt{R^2-\rho^2}}
-e^{-2i\left(k\sin^2{\theta\over2}+{G\over
2(ik_a+k)}\right)\sqrt{R^2-\rho^2}}\right]
\over{G\over2(ik_a+k)}+2k\sin^2{\theta\over2}}
\end{eqnarray*}
and the scattering length is
\begin{eqnarray*}
\label{qw4} &&f(0,0,k_a)=k_a\int\limits_0^R d\rho~\rho \left[1
-e^{-{G\over k_a}\sqrt{R^2-\rho^2}}\right]\\
&&={k_a R^2\over2}\left[1-{2\over
A^2}\left(1-(1+A)e^{-A}\right)\right],~~~A={GR\over k_a}
\end{eqnarray*}

The cross section looks as
\begin{eqnarray*}
\label{ff} && \sigma(k,k_s)=8\pi\int\limits_0^R dr~r\sin^2\left(
{G\over \sqrt{k_s^2+k^2}}\sqrt{R^2-r^2}\right)\\
&&={\pi R^2\over
B^2}\left[1+2B^2-\cos(2B)-2B\sin(2B)\right],~~~~~~B={GR\over\sqrt{k_s^2+k^2}}
\end{eqnarray*}

In  {\bf the quantum mean approximation for amplitudes} the
parameter $k_a$ is defined by the equation
\begin{eqnarray*}
\label{ff} && a=f(0,0,k_a)
\end{eqnarray*}
or
\begin{eqnarray*}
&& k_a\int\limits_0^R d\rho~\rho \left[1 -e^{-{G\over
2k_a}\sqrt{R^2-\rho^2}}\right]=G\int\limits_0^R dr~r^2 e^{-{G\over
2b}\left(R^2-r^2\right)}.
\end{eqnarray*}
The cross section is defined by the standard formula
\begin{eqnarray*}
\label{ff} && \sigma(k,k_a)=2\pi\int\limits_0^\pi
d\theta\sin\theta|f(k,k_a)|^2.
\end{eqnarray*}

In   {\bf the quantum mean approximation for cross section} the
parameter $k_s$ is defined by the equation
\begin{eqnarray*}
\label{ff} && 4\pi a^2=\sigma(0,k_s)
\end{eqnarray*}

In  {\bf the unitary approximation} the forward scattering amplitude
equals
\begin{eqnarray}
\label{up2} &&f(k,0)=(-k_c+ik)\int\limits_0^R d\rho~\rho\left[1-
e^{-{G\over k_c-ik}\sqrt{R^2-\rho^2}}\right]\\
&&={R^2\over2}(-k_c+ik)\left[1-{2\over
D^2}\left(1-(1+D)e^{-D}\right)\right],~~~~D={GR\over
k_c-ik}.\nonumber
\end{eqnarray}
The scattering length is
\begin{eqnarray}
\label{up2} &&{\cal A}(k_c)=f(0,0)={R^2\over2}(-k_c)\left[1-{2\over
D^2}\left(1-(1+D)e^{-D}\right)\right],~~~~D={GR\over k_c}.\nonumber
\end{eqnarray}
The imaginary part of the the amplitude reads
\begin{eqnarray}
\label{up2} &&{\cal I}(k_c)=\left.{{\rm Im}f(k,0)\over
k}\right|_{k=0}= R^2\left[{1\over2}-{3k_c^2\over G^2R^2}+e^{-{G\over
k_c}}\left[1+{3k_c\over GR}+{3k_c^2\over
G^2R^2}\right]\right]\nonumber\\
\end{eqnarray}
The parameter $k_c$ is defined by the equation
\begin{eqnarray}
\label{up2} &&4\pi {\cal A}^2(k_c)=4\pi{\cal I}(k_c)
\end{eqnarray}
Finally the cross section according to the optic theorem is
\begin{eqnarray}
\label{up2} &&\sigma(k)=4\pi{{\rm Im}f(k,0)\over k}
\end{eqnarray}

\vspace{1cm}

The comparison of the exact cross section with the approximations
are shown on Figures  1 and 2.

\subsection{Singular repulsing potential}

Let us calculate the scattering length and the cross section in the
case of scattering on the singular repulsing potential. In the
dimensionless variable  $r\to{r\over R}$ the hamiltonian reads
\begin{eqnarray}
\label{si1} && H={{\bf p}^2\over 2m}+{g\over R^2}\left({R^2\over{\bf
r}^2}\right)^N={1\over mR^2}\left[{{\bf p}^2\over 2}+{G\over
r^{2N}}\right],~~~~G=gm.
\end{eqnarray}
In the limit $N\to\infty$  the singular potential becomes the
potential of the rigid sphere
\begin{eqnarray}
\label{si2} && V(r)={g\over R^2}\left({R^2\over{\bf
r}^2}\right)^N\stackrel{N\to\infty}{\longrightarrow}\left\{\begin{array}{cc}
+\infty &r<R,\\
0&r>R.\\
\end{array}\right.
\end{eqnarray}

The perturbation approximation does not work in this case. The
scattering length is in the quantum mean approximation, according to
formula  (\ref{pqs4})
\begin{eqnarray}
\label{si3} && a(G)=G\int\limits_0^\infty{dr\over r^{2N-2}}
e^{-{G\over b}\int\limits_r^\infty{ds\over s^{2N-1}}} =\left({G\over
N-1}\right)^{{1\over2(N-1)}}f_a(N),\\
&&f_a(N)=b^{1-{1\over2(N-1)}}\Gamma\left(1-{1\over2(N-1)}\right).\nonumber
\end{eqnarray}

For a given singular potential the exact result for the scattering
length is known (see \cite{Flugge}):
\begin{eqnarray}
\label{si4} && a= \left({G\over
N-1}\right)^{{1\over2(N-1)}}f(N),~~~~f(N)= 2^{-{1\over
N-1}}{\Gamma\left(1-{1\over2(N-1)}\right)\over
\Gamma\left(1+{1\over2(N-1)}\right)}.
\end{eqnarray}
The behavior of the functions $f_a(N)$ for $b=1$ and $f_a(N)$ is
shown on Figure 3. One can see that in the case of large $N\gg1$
both the results coincide.

In the eiconal approximation the cross section equals
\begin{eqnarray}
\label{si5} && \sigma_e(k)= 8\pi\int\limits_0^\infty d\rho~\rho
\sin^2\left[{G\over k}\int\limits_0^\infty
{d\tau \over(\tau^2+\rho^2)^N}\right]\\
&&=2\pi\left({G\over
k}\right)^{{2\over2N-1}}\cdot\left[{\sqrt{\pi}~\Gamma\left(N-{1\over2}\right)\over
\Gamma(N)}\right]^{{2\over2N-1}}
\Gamma\left({2N-3\over2N-1}\right)\sin\left({\pi\over2}\cdot{2N+1\over2N-1}\right).\nonumber
\end{eqnarray}

For  $N\to\infty$ one has according (\ref{si4}) and (\ref{si5})
\begin{eqnarray}
\label{si6} && \sigma(k)\to\left\{\begin{array}{ll} 4\pi &
k\to0\\
&\\
2\pi  & k\to\infty\\
\end{array}\right.
\end{eqnarray}
In this limit $N\to\infty$  the dependence on momenta disappears in
the eiconal approximation.

\subsection{The Yukawa potential}

Let us calculate the scattering length and the cross section on the
Yukawa potential.

In the dimensionless variable $r\to{r\over\mu}$ the hamiltonian in
this case reads
\begin{eqnarray*}
\label{yu1} && H={{\bf p}^2\over 2m}+{g\over 4\pi}{e^{-\mu r}\over
r}={\mu^2\over m}\left[{{\bf p}^2\over2}+G{e^{-r}\over
r}\right],~~~G={gm\over4\pi \mu}
\end{eqnarray*}

\subsubsection{Quantum mean approximation}

The scattering length looks like
\begin{eqnarray}
\label{yu2} && a(b)= 2\int\limits_0^\infty dr~r e^{-r-{2G\over
b}e^{-r}}
\end{eqnarray}
The elastic cross section takes the form
\begin{eqnarray}
\label{yu3} &&\sigma_a(k,k_c)\approx 8\pi \int\limits_0^\infty
d\rho~\rho \sin^2\left({G\over\sqrt{k_c^2+k^2}}K_0(\rho)\right).
\end{eqnarray}
The parameter $k_c$ is defined by the condition that for the zeroth
momenta  two formulas (\ref{yu2}) and (\ref{yu3}) for the cross
section give the same result
\begin{eqnarray}
\label{yu4} && 4\pi a^2(b)=\sigma_a(0,k_c).
\end{eqnarray}

The results of calculations are shown on Figures 4 and 5.

\subsubsection{Unitary approximation}

The scattering amplitude in this approximation equals
\begin{eqnarray}
\label{yu5} &&f_a(k,\theta)= G\int\limits_0^\infty d\rho~\rho
J_0\left(k\rho\sin\theta\right)\\
&&\cdot\int\limits_{-\infty}^\infty dz~
e^{2ikz\sin^2{\theta\over2}}\cdot{e^{-\sqrt{z^2+\rho^2}}\over\sqrt{z^2+\rho^2}}\cdot
e^{-{iG\over ik_0+k}\int\limits_z^\infty ds~
{e^{-\sqrt{s^2+\rho^2}}\over\sqrt{z^2+\rho^2}}}.\nonumber
\end{eqnarray}

For the scattering length one gets
\begin{eqnarray}
\label{yu6} &&f_a(0,0)=G\int\limits_0^\infty d\rho~\rho
\int\limits_{-\infty}^\infty
dz~{e^{-\sqrt{z^2+\rho^2}}\over\sqrt{z^2+\rho^2}}\cdot e^{-{G\over
k_0}\int\limits_z^\infty ds~
{e^{-\sqrt{s^2+\rho^2}}\over\sqrt{z^2+\rho^2}}}\\
&&=k_0\int\limits_0^\infty d\rho~\rho \left[1- e^{-{2G\over
k_0}K_0(\rho)}\right].\nonumber
\end{eqnarray}
because
\begin{eqnarray*}
&& \int\limits_0^\infty ds
{e^{-\sqrt{s^2+\rho^2}}\over\sqrt{s^2+\rho^2}}= K_0(\rho).
\end{eqnarray*}
where $K_0(\rho)$ is the Bessel function of imaginary argument.-

The imaginary part of the scattering amplitude equals
\begin{eqnarray}
\label{yu7} &&{\rm Im}f_a(k,0)=-k\int\limits_0^\infty d\rho~\rho
\left[1- e^{-{2Gk_0\over k_0^2+k^2}K_0(\rho)}\cos\left({2Gk\over
k_0^2+k^2}K_0(\rho)\right)\right]\\
&&+k_0\int\limits_0^\infty d\rho~\rho~e^{-{2Gk_0\over
k_0^2+k^2}K_0(\rho)}\sin\left({2Gk\over
k_0^2+k^2}K_0(\rho)\right).\nonumber
\end{eqnarray}
For zero momenta $k=0$ one has
\begin{eqnarray} \label{yu8}
&&\left.{{\rm Im}f_a(k,0)\over k}\right|_{k=o}=-\int\limits_0^\infty
d\rho~\rho \left[1-\left(1+ {2G\over
k_0}K_o(\rho)\right)~e^{-{2G\over
k_0}K_0(\rho)}\right].\nonumber\\
\end{eqnarray}
The unitary condition for zero momenta $k=0$ gives the equation for
the parameter  $k_0$:
\begin{eqnarray*}
\label{yu9} &&\sigma(0)=4\pi\left(k_c\int\limits_0^\infty
d\rho~\rho\left[1- e^{-{2G\over
k_c}K_0(\rho)}\right]\right)^2\\
&&=4\pi\int\limits_0^\infty d\rho~\rho\left[1- e^{-{2G\over
k_c}K_0(\rho)}\left(1+{2G\over k_c}K_0(\rho)\right)\right]
\end{eqnarray*}

After calculation of the parameter  $k_c=k_c(G)$ as a function of
the coupling constant one can compute the cross section
\begin{eqnarray}
\label{yu10} &&\sigma(k)\\
&&=4\pi\int\limits_0^\infty d\rho~\rho\left\{1- e^{-{2Gk_c\over
k_c^2+k^2}K_0(\rho)}\left[\cos\left({2Gk\over
k_c^2+k^2}K_0(\rho)\right)+ {k_c\over k}\sin\left({2Gk\over
k_c^2+k^2}K_0(\rho)\right)\right]\right\}.\nonumber
\end{eqnarray}

On Figures 6 and 7 the results of calculations are demonstrated.

\section{Conclusion}

The representation of the elastic scattering amplitude in the form
of the path integral is obtained by using the stationary
Schroedinger equation.

The methods of evaluation of the path integrals are based on the
uncertainty correlation for a free motion.

Formulas for the scattering lengths and cross sections for any
coupling constants are simple enough so that qualitative and
semiquantitative estimations can be obtained without great efforts.
The examined examples show the effectiveness of the proposed
methods.

Generally speaking, other known and unknown methods of calculating
of path integrals can be worked out and applied, for example,
variation methods and method of the Gaussian equivalent
representation (see, for example, \cite{Efim1,Efim2,Rosen,Ros}) and
so on). All these problems require further investigations.

In conclusion I wish to thank V.S.Melezhik for many helpful
discussions.

\newpage

\begin{figure}[ht]\
\begin{center}
\epsfig{figure=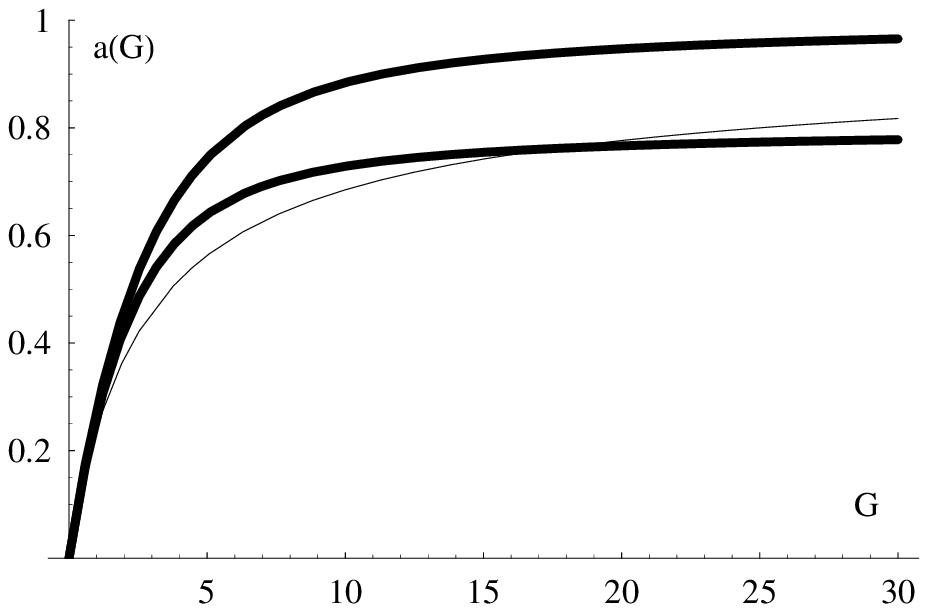,width=10cm}
\end{center}
\caption{The scattering length for the right angled potential as a
function of the coupling constants $G$. Boldface lines - the quantum
mean approximation. The upper line - $b=1$, lower line - $b=0.8$.
Thin line - exact result.}
\end{figure}

\begin{figure}[ht]
\begin{center}
\epsfig{figure=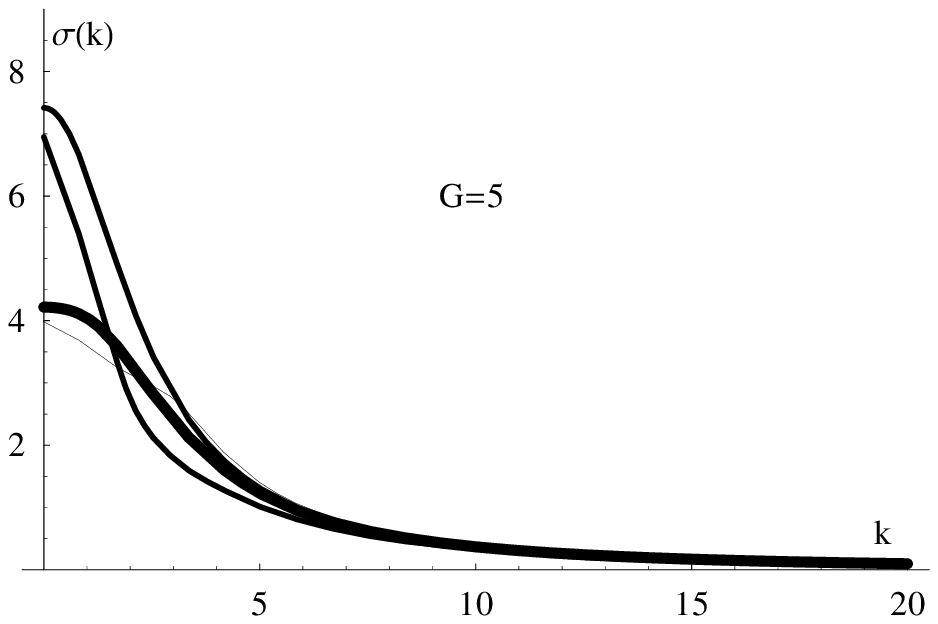,width=8cm}\\
\epsfig{figure=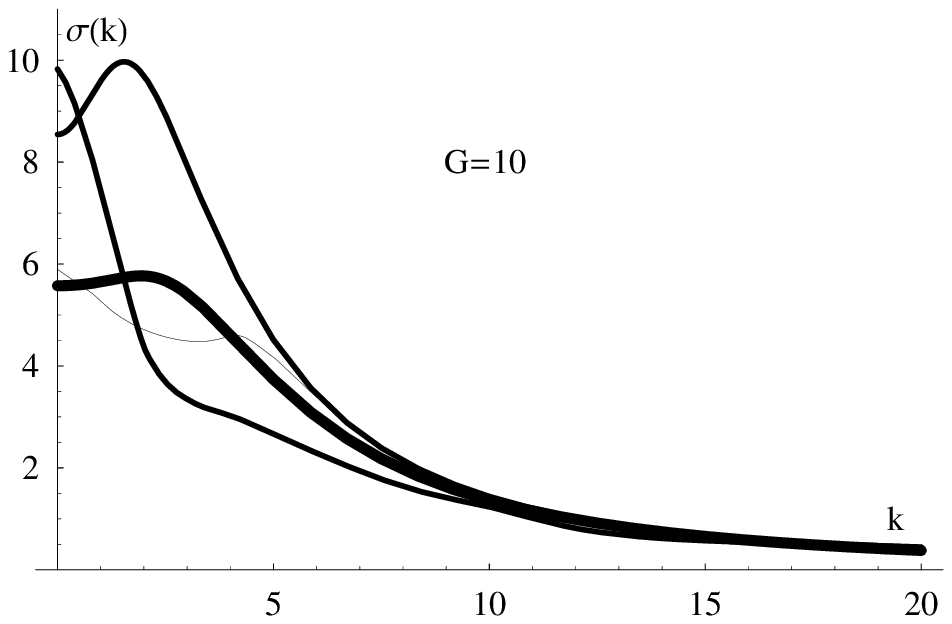,width=8cm}\\
\epsfig{figure=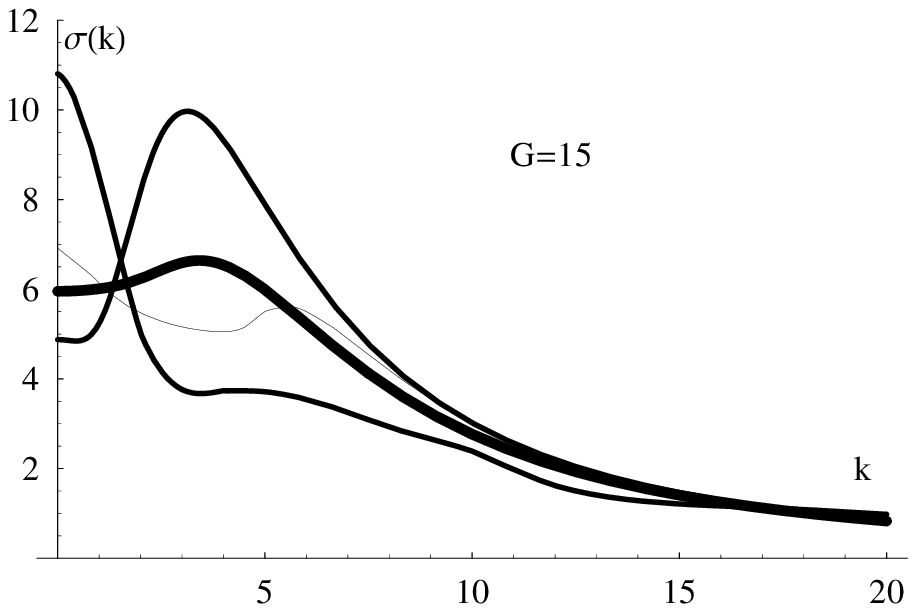,width=8cm}\\
\end{center}
\caption{The cross section $\sigma(k)$ for the right angled
potential for the coupling constants  $G=5$, $G=10$ and $G=15$ .
Boldface lines - the quantum mean approximation. The upper line -
$b=1$, lower line - $b=0.8$. Thin line - exact result.}
\end{figure}

\begin{figure}[ht]\
\begin{center}
\epsfig{figure=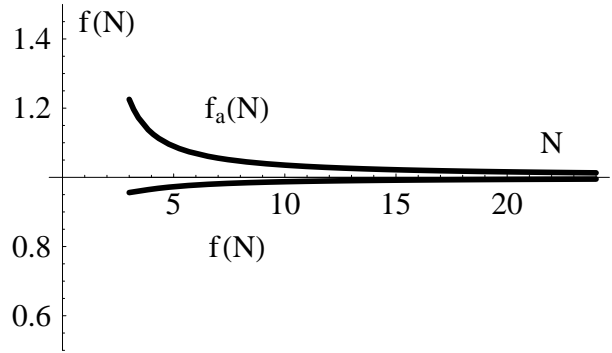,width=8cm}
\end{center}
\caption{ The scattering length for the
singular potential as a function of  the coupling constant $G$. The
upper line - quantum mean approximation, lower line - exact result.}
\end{figure}

\begin{figure}[ht]
\begin{center}
\epsfig{figure=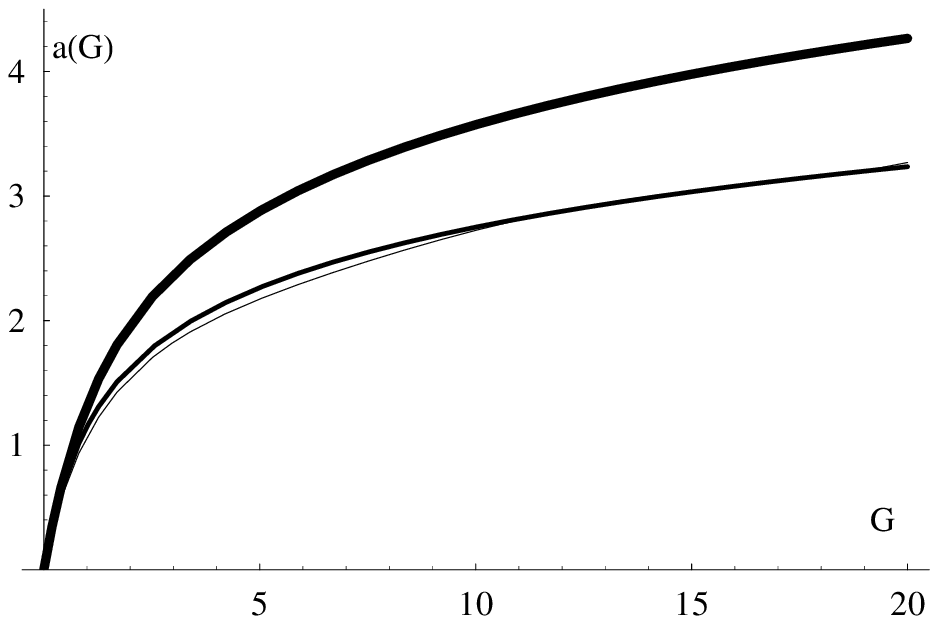,width=12cm}
\end{center}
\caption{Quantum mean approximation. The scattering length for the
Yukawa potential as a function of  the coupling constant $G$.
Boldface lines - the approximation for $b=1$, thiner lines  -
$b=0.7$. Thin line - numerical result. } \label{fig5}
\end{figure}

\begin{figure}[ht]
\begin{center}
\epsfig{figure=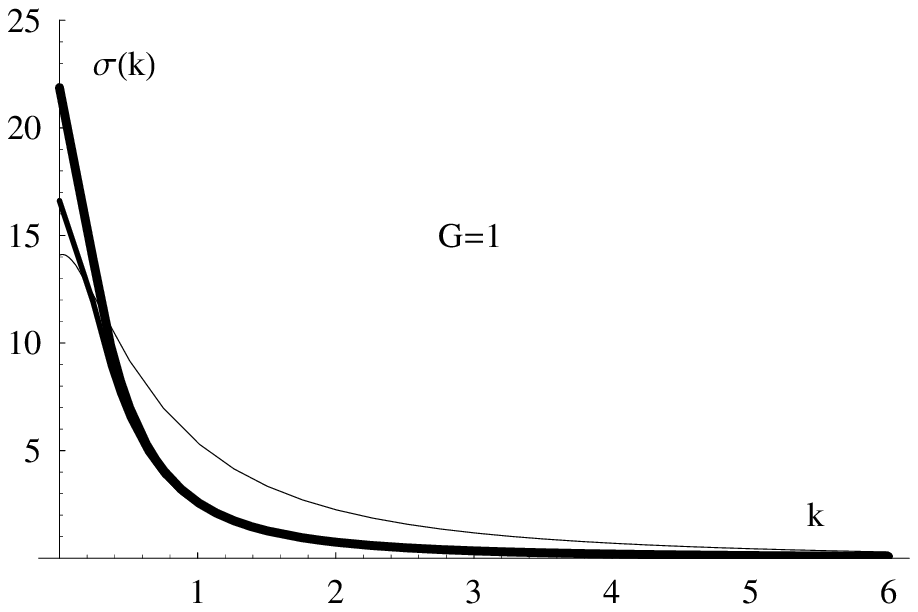,width=8cm}\\
\epsfig{figure=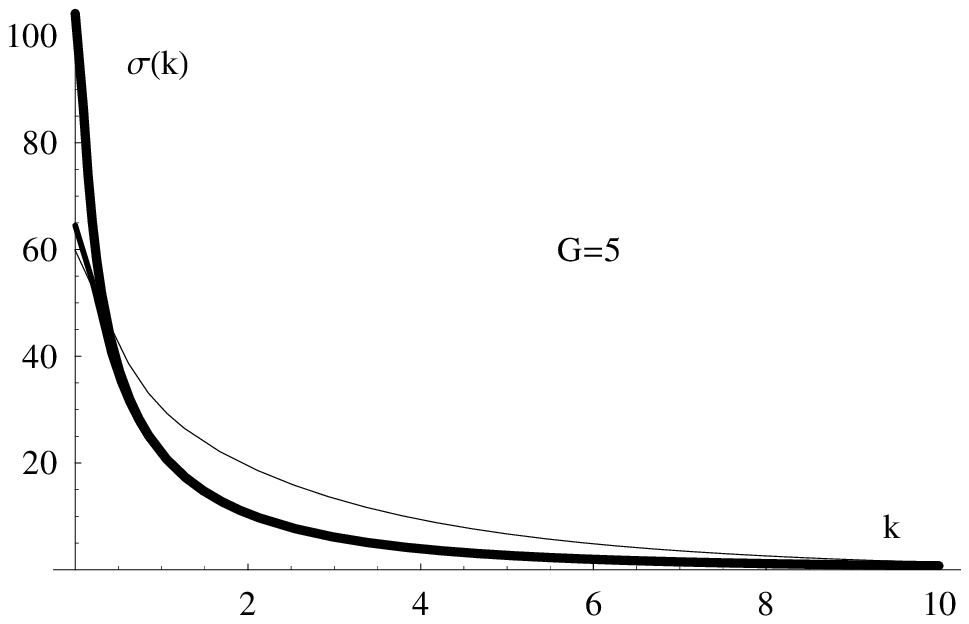,width=8cm}\\
\epsfig{figure=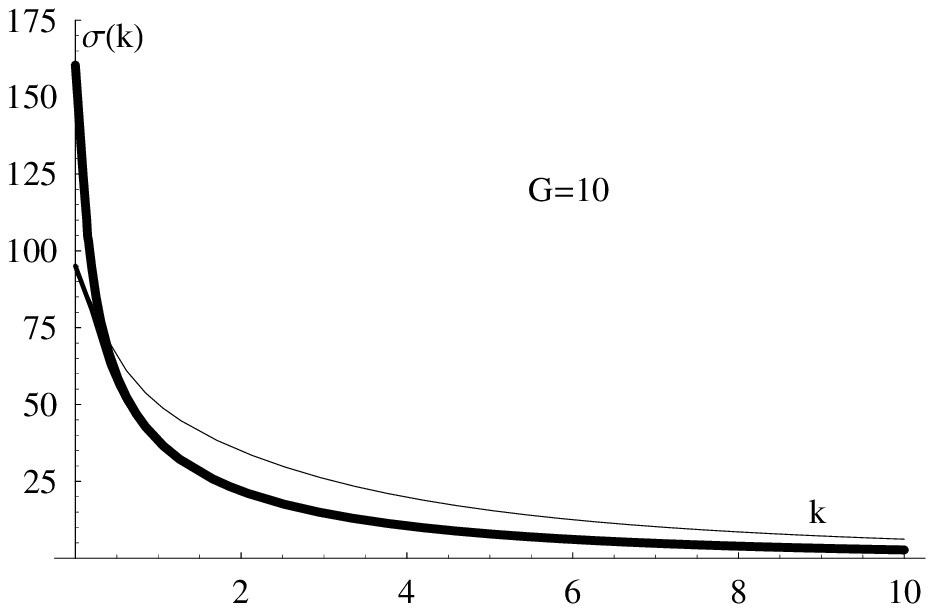,width=8cm}
\end{center}\label{sech-4}
\caption{Quantum mean approximation. The cross section $\sigma(k)$
for the Yukawa potential for the coupling constants $G=5$, $G=10$
and $G=15$ . Boldface lines - the approximation, thin line -
numerical result. }
\end{figure}

\begin{figure}[ht]
\begin{center}
\epsfig{figure=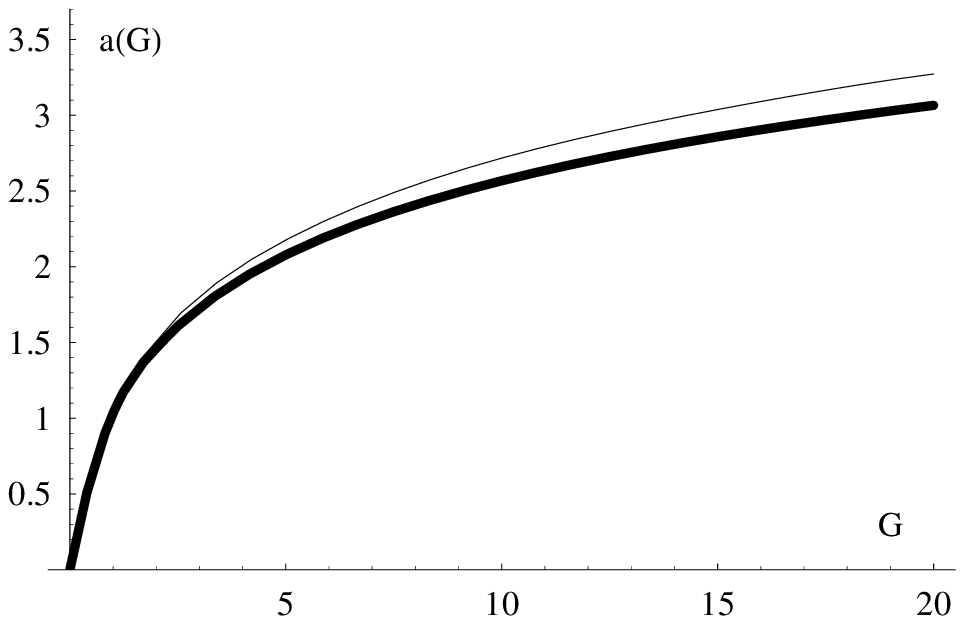,width=12cm}
\end{center}\label{sech-5}
\caption{Unitary approximation. The scattering length for the Yukawa
potential as a function of the coupling constant $G$. Boldface lines
- the approximation for $b=1$, thiner lines  for $b=0.7$. Thin line
- numerical result. }
\end{figure}

\begin{figure}[ht]
\begin{center}
\epsfig{figure=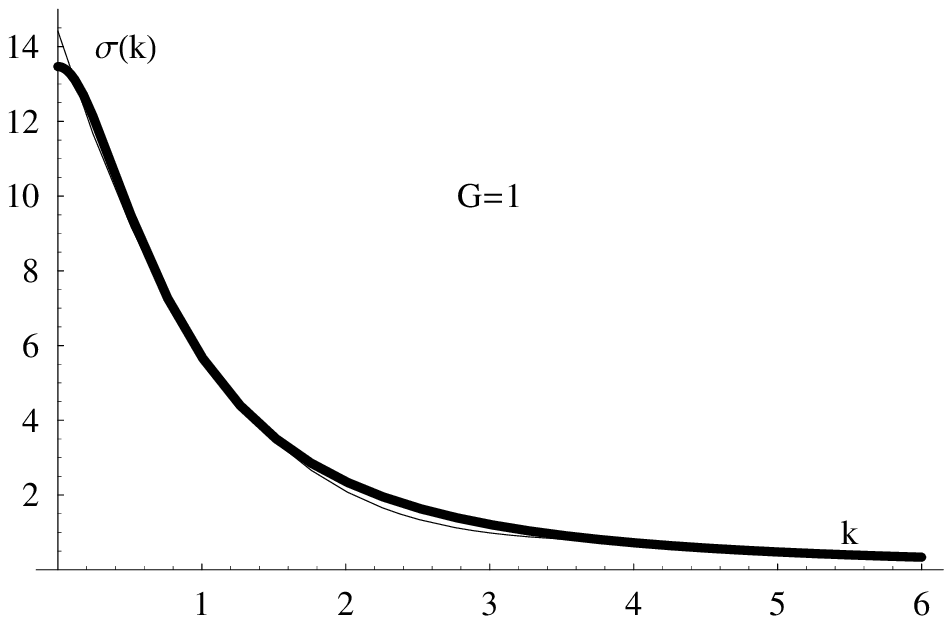,width=8cm}\\
\epsfig{figure=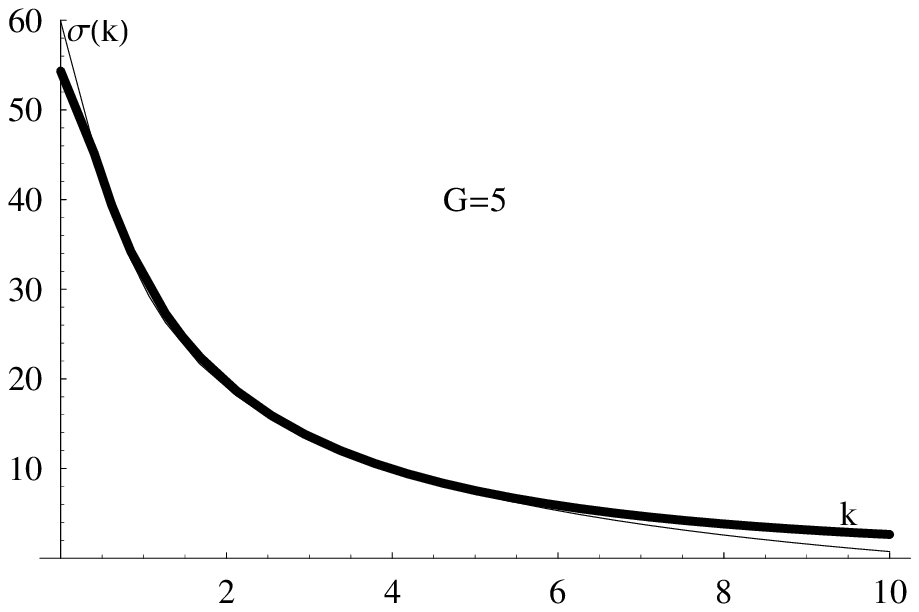,width=8cm}\\
\epsfig{figure=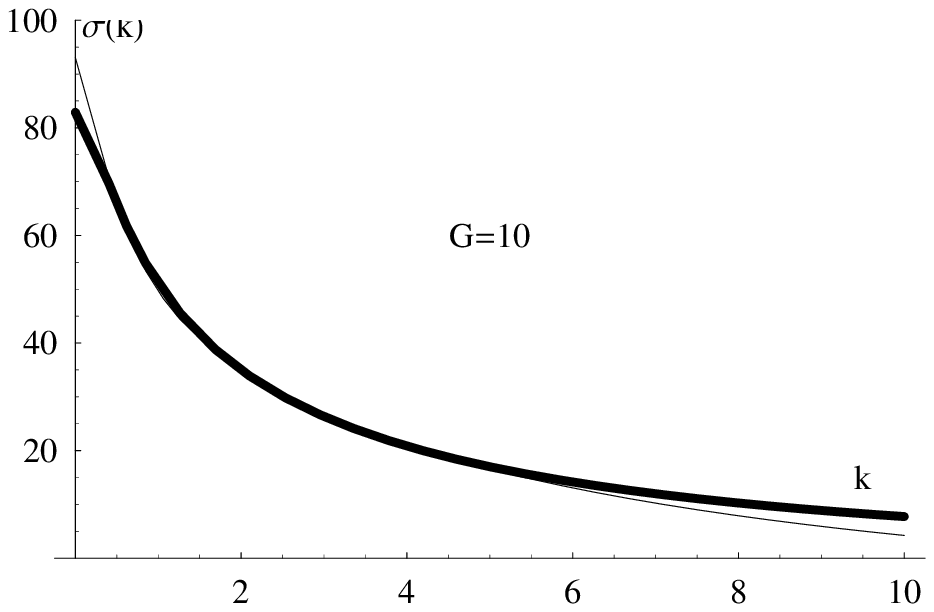,width=8cm}
\end{center}\label{sech-6}
\caption{Unitary approximation. The cross section $\sigma(k)$ for
the Yukawa potential for the coupling constants $G=5$, $G=10$ and
$G=15$ . Boldface lines - the approximation, thin line - numerical
result.}
\end{figure}

\end{document}